\documentclass[aps,twocolumn]{revtex4}
\usepackage[dvips]{graphicx}
\newcommand{\sect}[1]{\setcounter{equation}{0}\section{#1}}

\begin{document}
\title{Relativistic kinetics and power-law tailed distributions}
\author{G. Kaniadakis}
\email{giorgio.kaniadakis@polito.it}
\affiliation{Dipartimento di Fisica, Politecnico di Torino, \\
Corso Duca degli Abruzzi 24, 10129 Torino, Italy}
\date{\today}

\begin {abstract}

 The present paper is devoted to the relativistic statistical theory, introduced in Phys. Rev. E {\bf 66} (2002) 056125 and Phys. Rev. E {\bf 72} (2005) 036108, predicting the particle distribution function $p(E)= \exp_{\kappa} (-\beta[E-\mu])$ with $\exp_{\kappa}(x)=(\sqrt{1+ \kappa^2 x^2}+\kappa x)^{1/\kappa}$, and $\kappa^2<1$. This, experimentally observed, relativistic distribution, at low energies behaves as the exponential, Maxwell-Boltzmann classical distribution, while at high energies presents power law tails. Here, we obtain the evolution equation, conducting asymptotically to the above distribution, by using a new deductive procedure, starting from the relativistic BBGKY hierarchy and by employing the relativistic molecular chaos hypothesis.

\end {abstract}


\maketitle

\sect{Introduction}

In the last few decades, the power-tailed statistical distributions have been observed in a variety of physical, natural or artificial systems. In \cite{EPJB2009} one can find an updated discussion of power-law tailed distributions and an extensive list of systems where that distributions have been empirically observed.

For instance the cosmic ray spectrum, $f_i\propto \chi(\beta E_i-\beta \mu)$, obeys the Boltzmann law of classical statistical mechanics i.e. $\chi(x)
{\atop\stackrel{\textstyle\sim}{\scriptstyle x\rightarrow \,0}}
\exp(-x)$ at low energies, while at high energies this spectrum
presents power law fat tails i.e $\chi(x)
{\atop\stackrel{\textstyle\sim}{\scriptstyle x\rightarrow
\,+\infty}} x^{-a}$. The cosmic rays problem  was
approached for the first time, by using a different
distribution from the Boltzmann one, in 1968. In his proposal Vasyliunas
heuristically identified the function $\chi(x)$ with the Student
distribution function which presents power-law tails
\cite{EPJB2009}.

Currently, there is an intense debate regarding the physical origin of the experimentally observed non-Boltzmannian distributions. Recently, after noting that the power-law tails are placed in the high energy region, where the particles are relativistic, the question has been posed whether the solution of the problem, i.e. the theoretic determination of the function $\chi(x)$ and consequently of the related distribution and
entropy, can be explained by invoking the basic principles of special relativity. Few years ago in refs. \cite{PRE02,PRE05} has been obtained within the special relativity the following simple expression for the function $\chi(x)$, i.e. $\chi(x)=\exp_{\kappa}(x)$ with
\begin{equation}
\exp_{\kappa}(x)=(\sqrt{1+ \kappa^2 x^2}+\kappa x)^{1/\kappa}.
\label{A1}
\end{equation}
The parameter $\kappa^2<1$ is the reciprocal of light speed, in a dimensionless form, while $1/\kappa^2$ is the particle rest energy. In the classical limit $\kappa \rightarrow 0$, $\exp_{\kappa}(x)$ reduces to the ordinary exponential function, defining the Boltzmann factor of classical statistical mechanics \cite{PhA01,PLA01,EPJA2009}.

In the last few years various authors have considered the
foundations of this statistical theory, e.g. the H-theorem and the
molecular chaos hypothesis \cite{Silva06A,Silva06B}, the
thermodynamic stability \cite{Wada1,Wada2}, the Lesche stability
\cite{KSPA04,AKSJPA04,Naudts1,Naudts2}, the Legendre structure of
the ensued thermodynamics \cite{ScarfoneWada,Yamano}, the geometrical structure of the theory \cite{Pistone}, etc. On the other hand
specific applications to physical systems have been considered, e.g.
the cosmic rays \cite{PRE02}, relativistic \cite{GuoRelativistic}
and classical \cite{GuoClassic} plasmas  in presence of external
electromagnetic fields, the relaxation in relativistic plasmas under
wave-particle interactions \cite{Lapenta,Lapenta2009}, anomalous diffusion \cite{WadaScarfone2009,Wada2010}, kinetics
of interacting atoms and photons \cite{Rossani}, particle kinetics in the presence of temperature gradients \cite{GuoDuoTgradient},  particle systems in external conservative force fields \cite{Silva2008}, stellar distributions in astrophysics \cite{Carvalho,Carvalho2}, quark-gluon
plasma formation \cite{Tewel}, quantum hadrodynamics models
\cite{Pereira}, the fracture propagation \cite{Fracture}, etc. Other applications regard
dynamical systems at the edge of chaos \cite{Corradu,Tonelli,Celikoglu},
fractal systems \cite{Olemskoi}, the random matrix theory
\cite{AbulMagd,AbulMagd2009}, the error theory \cite{WadaSuyari06}, the game
theory \cite{Topsoe}, the information theory \cite{WadaSuyari07},
etc. Also applications to economic systems have been considered e.g.
to study the personal income distribution
\cite{Clementi,Clementi2008,Clementi2009}, to model deterministic heterogeneity in tastes and product differentiation \cite{Rajaon,Rajaon2008} etc.

In the present contribution we reconsider critically the kinetic foundations of the statistical theory based on the function $\exp_{\kappa}(x)$. Our main goal is to show that the standard principles of ordinary relativistic statistical physics, {\it i.e.} kinetic equation, H-theorem, Molecular Chaos Hypothesis (MCH), conduct unambiguously to the relativistic generalization of the classical Boltzmann entropy and Maxwell-Boltzmann distribution.

\sect{Kinetic Equation}

Let us consider the most general relativistic equation imposing the particle conservation during the evolution of a many body system. That equation is the first equation of the BBGKY hierarchy i.e.
\begin{equation}
p^{\,\nu}\partial_{\nu}f-m F^{\nu}\frac{\partial f}{\partial
p^{\,\nu}}= \!\int \!
\frac{d^3p'}{{p'}^{0}}\frac{d^3p_1}{p_1^{\,0}}\frac{d^3p'_1}{{p'}_{\!\!1}^{0}}
\,\,G \,\left[f'\!\otimes \!f'_1\!-\! f\!\otimes \! f_1\right],
\label{B1}
\end{equation}
and describes, through the one-particle correlation function or distribution function $f=f(x,p)$, a relativistic
particle system in presence of an external force field. The streaming term as well as the Lorentz invariant integrations in the collision integral, have the standard forms forms of the relativistic kinetic theory \cite{Degroot,Cercignani}. The two particle correlation function $F(f,f_1)$ \cite{KersonHuang}, here denoted by $f\!\otimes \! f_1$, at the moment remains an unknown function, which will be determined in the following by employing the relativistic MCH. We recall that in classical kinetics, the two-particle correlation function, according to the MCH, is assumed to be, $f\!\otimes \! f_1= f\,f_1$.

Following standard lines of kinetic theory, we note that in
stationary conditions, the collision integral in Eq. (\ref{B1})
vanishes and then it follows
\begin{eqnarray}
f\!\otimes \! f_1= f'\!\otimes \! f'_1  \ \ . \label{B2}
\end{eqnarray}
More in general it holds $L(f\!\otimes \! f_1)= L(f'\!\otimes
\! f'_1)$, being $L(x)$ an arbitrary function, and this relationship
expresses a conservation law for the particle system. On the other
hand a conservation law has form
\begin{equation}
\Lambda(f)+ \Lambda(f_{1}) =\Lambda(f^\prime) +\Lambda(f^\prime_{1})
\ , \ \ \label{B3}
\end{equation}
$\Lambda (f)$ being the
collision invariant of the system. Therefore we can
pose $L(f\!\otimes\!f_1)= \Lambda(f)+\Lambda(f_1)$.

From the definition of the correlation
function and taking into account the indistinguishability of
the particles, it follows that $f\!\otimes \!1=1\otimes \!f=f$ and this implies that $L(f)=\Lambda(f)$ and $\Lambda(1)=0$. Therefore the function $\Lambda(f)$ permit us to determine univocally the correlation function $f\!\otimes\!f_1$ through the equation
\begin{eqnarray}
\Lambda(f\!\otimes\!f_1)= \Lambda(f)+\Lambda(f_1) \ \ . \label{B4}
\end{eqnarray}

The composition law $f\otimes f_1$ has the properties:

\noindent (i) $(f\otimes
f_1)\otimes f_2=f\otimes (f_1\otimes f_2)$, i.e. it is associative,

\noindent (ii)  $f\otimes 1=1\otimes f=f$, i.e. it admits as neutral element the unity,

\noindent (iii) $f\otimes f_1=f_1 \otimes f$, i.e. it is commutative.

The above three properties are the same of the ordinary product. In order to become $\otimes$ a generalized product isomorphic to the ordinary product, it is necessary and sufficient to require the property

\noindent (iv) $f\otimes(1/f)=(1/f)\otimes f=1$, i.e. the inverse element of $f$ is $1/f$. Then, the real, positive, probability distribution functions,
form an Abelian group.

The function $\Lambda(f)$, through Eq. (\ref{B4}), defines
univocally the two-particle correlation function $f\otimes f_1$
which is a generalized product having the same algebraic properties
of the ordinary product.

It is remarkable that the property (iv) imposes to the function $\Lambda(f)$ the condition
\begin{eqnarray}
\Lambda(1/f)= -\Lambda(f) \ \ . \label{B5}
\end{eqnarray}

In relativistic kinetics, the collision invariant
$\Lambda(f)$, unless an additive constant, is proportional to the microscopic relativistic invariant $I$. Then we can pose
\begin{equation}
\Lambda(f)=- \beta I +\beta \mu \ ,
\label{B6}
\end{equation}
$\beta $ and $\beta \mu$ being two arbitrary constants. In presence of an external electromagnetic field $A^{\nu}$, the more general
microscopic relativistic invariant $I$, has a form proportional to
\begin{equation}
I=\left(p^{\nu}+q A^{\nu}\!/c \right)\,U_{\nu}-mc^2 \ ,
\label{B7}
\end{equation}
$U_{\nu}$ being the hydrodynamic four-vector velocity with
$U^{\nu}U_{\nu}=c^2$ \cite{Degroot}.

Finally, after inversion of Eq. (\ref{B6}), it obtains
the stationary distribution as follows
\begin{equation}
f=\Lambda^{-1}\big(-\beta\,I+\beta \,\mu \,\big) \ . \label{B8}
\end{equation}

It is remarkable to note that Eq. (\ref{B6}), after being written in the form $- \Lambda (f)- \beta \, I
+ \beta \mu =0$, follows from the variational equation
\begin{equation}
\frac{\delta}{\delta f}\,\,\Bigg[ \, S \,-\, \beta \!\!\int \!\!d^3p \,\,I\,f\, \,+\, \beta
\mu \!\!\int\!\! d^3p \,\, f \,\Bigg]=0 \ ,  \ \ \
\label{B9}
\end{equation}
where the functional $S$, unless an arbitrary additive constant, is given by
\begin{equation}
S= \int \!\!d^3p \, \, \, \sigma (f) \, df   \ \ ,
\label{B10}
\end{equation}
where the entropy density $\sigma (f)$ is given by
\begin{equation}
 \Lambda (f)= -\frac{\partial \sigma (f)}{ \partial f} \ \ .
\label{B11}
\end{equation}

The variational equation (\ref{B9}), defining univocally the distribution (\ref{B8}), represents  the Maximum Entropy Principle (MEP). The constants $\beta$ and $\beta \mu$ are the Lagrange multipliers while the functional $S$, defined through Eq. (\ref{B10}), is the system entropy. We stress that both the stationary distribution (\ref{B8}) and the entropy  (\ref{B10}) depend on the form of the function $\Lambda (f)$. This function is related to the two-particle correlation function through (\ref{B4}), and its expression will be obtained in the next sections.

\sect{The H-theorem}

In the standard relativistic kinetics it is well
known from the H-theorem that the production of entropy is never
negative and in equilibrium conditions there is no entropy
production. In the following we will demonstrate the H-theorem
for the system governed by the kinetic equation (\ref{B1}). We define the four-vector
entropy $S^{\nu}=(S^{0},\mbox{\boldmath $S$})$ as follows
\begin{equation}
S^{\nu}= \int \frac{d^3p}{p^{0}}\,p^{\nu}\, \, \sigma (f)
\ , \label{H1}
\end{equation}
and note that $S^{0}=S$ coincides with the scalar entropy defined
in Eq. (\ref{B10}), while $\mbox{\boldmath $S$}$ is the entropy flow. The identity
$d^3p/p^0=d^4p \,\, 2\,
\theta(p^0)\,\delta(p^{\mu}p_{\mu}-m^2c^2)$, permits us to write the four-vector
$S^{\nu}$ in the form
\begin{equation}
S^{\nu}= \int
d^4p\,\,2\,\theta(p^{0})\,\delta(p^{\mu}p_{\mu}-m^2c^2)\,\,p^{\nu}\,
\,\sigma (f) \ \ . \label{H2}
\end{equation}
In this expression $d^4p$ is a scalar because the Jacobian of the
Lorentz transformation is equal to unit. Then since $p^{\nu}$
transforms as a four-vector, we can conclude that $S^{\nu}$
transforms as a four-vector.

In order to calculate the entropy production $\partial_{\nu}S^{\nu}$ we start from the definition of $S^{\nu}$ given by (\ref{H1}), the evolution equation
(\ref{B1}) and take into account the relationship $\partial_{\nu} \, \sigma (f)=[\,\partial \,\sigma (f)/\,\partial f \,] \, \partial_{\nu}f=-\Lambda (f) \,\partial_{\nu}\,f$. We obtain
\begin{eqnarray}
\partial_{\nu}S^{\nu}=  &&\!\!\!\!\!-  \int
\frac{d^3p}{p^{0}}\,\ln_{\kappa}(g)\,\,p^{\nu}\,\partial_{\nu}f \nonumber \\
=&&\!\!\!\!\!-\int
\frac{d^3p}{p^{0}}\frac{d^3p'}{{p'}^{0}}\frac{d^3p_1}{p_1^{\,0}}\frac{d^3p'_1}{{p'}_{\!\!1}^{0}}
\,\,G \, \left[f'\!\otimes \!f'_1\!-\! f\!\otimes \! f_1\right] \, \Lambda (f) \nonumber \\
&&\!\!\!\!\!-  \,m \int
\frac{d^3p}{p^{0}}\,\Lambda (f)\,\,F^{\nu}\frac{\partial
f}{\partial p^{\,\nu}} \ \ . \ \ \ \label{H3}
\end{eqnarray}
Since the Lorentz force $F^{\nu}$ has the properties
$p^{\nu}F_{\nu}=0$ and $\partial F^{\nu}/\partial p^{\nu}=0$ the
last term in the above equation involving $F^{\nu}$ is equal to
zero \cite{Degroot}.
Given the particular symmetry of the non vanishing integral in Eq. (\ref{H3})
we can write the entropy production as follows
\begin{eqnarray}
\partial_{\nu}S^{\nu}=\!\!\!\!\!&&-\frac{1}{4}\, \int \frac{d^3p}{p^{0}}
\frac{d^3p'}{{p'}^{0}}\frac{d^3p_1}{p_1^{\,0}}\frac{d^3p'_1}{{p'}_{\!\!1}^{0}}
\,\,G \, \left[f'\!\otimes \!f'_1\!-\! f\!\otimes \! f_1\right] \nonumber \\
&&\times \, [\Lambda(f) + \Lambda(f_1)-\Lambda(f')
- \Lambda(f'_1) ] \ \ . \ \ \ \ \ \label{H5}
\end{eqnarray}

Finally, we set this equation in the form

\begin{eqnarray}
\partial_{\nu}S^{\nu}=\!\!\!\!\!&&\frac{1}{4}\, \int \frac{d^3p}{p^{0}}
\frac{d^3p'}{{p'}^{0}}\frac{d^3p_1}{p_1^{\,0}}\frac{d^3p'_1}{{p'}_{\!\!1}^{0}}
\,\,G \,  \nonumber \\
&&\times \left[f'\!\otimes \!f'_1\!-\! f\!\otimes \! f_1\right] [ \Lambda(f'\!\otimes \!f'_1) - \Lambda(f\!\otimes \! f_1) ] \ \ . \ \ \ \ \ \ \ \label{H5}
\end{eqnarray}

After imposing that $\Lambda(h)$ is an increasing function, it results
$[h_1-h_2]\,[\Lambda(h_1)-\Lambda(h_2)]\geq 0$
$\forall h_1,h_2$ and then we can conclude that
\begin{equation}
\partial_{\nu}S^{\nu}\geq 0 \ \ . \label{H7}
\end{equation}
This last relation is the local formulation of the relativistic
H-theorem which represents the second law of the thermodynamics
for the system governed by the evolution equation (\ref{B1}).

In the next section we will determine the form of the function $\Lambda (f)$ by employing the relativistic MCH.

\sect{Relativistic Statistical Theory}

In classical statistical mechanics, it is well known  that the MCH, imposing for the two-particle correlation function the classical form $f\otimes
f_1=f\,f_1$, implies according to Eq. (\ref{B4}), that $\Lambda(f)=\ln(f)$. As a consequence the expression of the entropy (\ref{B10}), unless an additive constant, simplifies as
\begin{equation}
S= - \,\gamma\! \int \!d^3p \,\, f \,\Lambda (f/\epsilon) \ ,
\label{C1}
\end{equation}
with $\gamma=1$ and $\epsilon=e$. It is worth stressing that the
latter simple form of the entropy, defined as the mean value of
$-\Lambda (f/\epsilon)$, is enforced exclusively by the MCH, and represents a very important feature of the ordinary statistical theory.

In order to better point out the role of the MCH, in the
construction of the ordinary statistical theory, we equal the expression (\ref{B10}), defining the entropy, with its expression as given in Eq. (\ref{C1}) accounting the MCH, and obtain the following differential-functional equation
\begin{eqnarray}
\frac{d}{df}\,\,f\,\Lambda(f)=\frac{1}{\gamma} \,\,\Lambda(\epsilon
f) \ \ . \label{C2}
\end{eqnarray}

The latter equation, imposed by the MCH, can be viewed as the starting point of an alternative approach to obtain the ordinary statistical theory.
Indeed, as first step, we obtain the solution $\Lambda(f)=\ln(f)$, $\gamma=1$, $\epsilon=e$ of Eq. (\ref{C2}), obeying the condition (\ref{B5}). As second step, through Eq. (\ref{B4}), we obtain the expression of the correlation function $f\!\otimes \! f_1= f\,f_1$ appearing in the evolution equation (\ref{B1}). As third step, by employing Eq. (\ref{C1}), we obtain the Boltzmman entropy $S= - \int d^3p \,\, f \,\ln\,(f/e)$. As fourth and last step, through Eq. (\ref{B8}), we obtain the Maxwell distribution $f=\exp\big(-\beta\,I+\beta \,\mu \,\big)$.

It is important to emphasize that Eq. (\ref{C2}), already known in the literature, is enforced exclusively by the MCH. In the light of this, spontaneously emerges the question if  Eq. (\ref{C2})
with the condition (\ref{B5}) admits other solutions, beside the
aforementioned classical solution. In other words we pose the question if there exist other expressions of the $\Lambda (f)$, apart the classical ones, preserving the
important feature (\ref{C2}) of the ordinary statistical mechanics,
enforced by the MCH. Interestingly, Eqs. (\ref{C2}) and (\ref{B5}) admit another, more general solution, which is unique and contains as limiting,
particular case, the classical solution. The prediction of this new solution represents a very strong argument in favor of the present approach.

The general solution of Eq. (\ref{C2}) is defined up to an arbitrary multiplicative constant which can be fixed by using the
normalization condition $\Lambda'(1)=1$ yielding
\begin{eqnarray}
\gamma=\Lambda(\epsilon) \ \  . \ \ \ \label{C3}
\end{eqnarray}
Consequently the general solution, depends only on one free parameter, in the following indicated with $\kappa$, and defines a generalized
logarithm i.e.
\begin{eqnarray}
\Lambda(x)=\ln_{\kappa}(x) \ \  . \ \ \ \label{C4}
\end{eqnarray}
After solving
Eq.(\ref{C2}) with the condition (\ref{B5}), one obtains \cite{PRE02} the following simple expressions for the generalized logarithm and its inverse
function i.e. the generalized exponential
$\Lambda^{-1}(x)=\exp_{\kappa}(x)$:
\begin{eqnarray}
&&\ln_{\kappa}(x)= \frac{x^{\kappa}-x^{-\kappa}}{2\kappa} \ ,
\label{C4} \\
&&\exp_{\kappa}(x)=\left(\sqrt{1+\kappa^2 x^2}+\kappa
x\right)^{1/\kappa}  , \ \ \ \ \ \ \ \label{C6}
\end{eqnarray}
with $\kappa^2<1$. The constants $\gamma$ and $\epsilon$, depend
on the parameter $\kappa$ and are given by
\begin{eqnarray}
\gamma=\frac{1}{\sqrt{1-\kappa^2}} \ ,  \ \ \ \ \ \ \ \epsilon=\exp_{\kappa}(\gamma)  . \ \ \ \ \ \ \ \label{C7}
\end{eqnarray}
The above generalized logarithm and exponential are
one-parameter continuous deformations of the ordinary functions
which recovers in the $\kappa \rightarrow 0$ limit.

The entropy (\ref{C1}) associated to the new solution becomes
\begin{eqnarray}
S=- \,\gamma\int d^3p \,\,\, f \,\ln_{\kappa}(f/\epsilon) \ ,
\label{C8}
\end{eqnarray}
and can be written explicitly as
\begin{eqnarray}
S= \frac{1}{2\kappa} \int
d^3p \,\,\bigg(\frac{f^{1-\kappa}}{1-\kappa} -
\frac{f^{1+\kappa}}{1+\kappa}\bigg) \ , \label{C9}
\end{eqnarray}
while the stationary distribution (\ref{B8}) assumes the form
\begin{eqnarray}
f=\exp_{\kappa}\!\big(\!-\beta\,I+\beta\,\mu \,\big) \ . \ \ \ \ \ \
\ \ \ \label{C10}
\end{eqnarray}
The latter entropy and distribution, in the $\kappa \rightarrow 0$
limit reduce to the classical Boltzmann entropy and Maxwell-Boltzmann distribution respectively.

The distribution (\ref{C10}) in the global rest frame where
$U_{\nu}=(c,0,0,0)$ and in absence of external forces i.e.
$A^{\nu}=0$, simplifies as
\begin{eqnarray}
f=\exp_{\kappa}\left(- \beta\,E+\beta\,\mu \right) \ . \ \ \ \ \ \
\ \ \ \label{C11}
\end{eqnarray}
$E$ being the relativistic kinetic
energy. This distribution at low energies ($E \rightarrow 0$)
reduces to the classical Maxwell-Boltzmann distribution i.e. $f\approx
\exp\left(- \beta\,E-\beta\,\mu \right)$, while at relativistic
energies ($E \rightarrow +\infty$) presents power-law tails
$f\propto E^{-1/\kappa}$, in accordance with the experimental
evidence.

After posing $\Lambda(f)=\ln_{\kappa}(f)$, in  Eq. (\ref{B4}), the relativistic two-particle correlation function assumes the form
\begin{equation}
f\otimes f_1 =\exp_{\kappa} \left( \,\ln_{\kappa}\! f  + \ln_{\kappa}\!f_1 \, \right )  \ ,  \label{C12}
\end{equation}
which in the $\kappa \rightarrow 0$ classical limit, reduces to  $ff_1$.

It is easy to verify that the distribution (\ref{C10}) is the stationary solution of the evolution equation (\ref{B1}). First we observe that in stationary conditions, both the steaming term (i.e. the left hand side) as well as the collision integral (i.e. the right hand side) of Eq. (\ref{B1}), vanish.

For the collision integral to vanish, it is necessary and sufficient to pose $f'\otimes f'_1-f\otimes f_1=0$. The latter equation, after taking into account Eqs. (\ref{C11}) and (\ref{C12}), produces the relationships $E' + E'_1=E + E_1$ and $\mu' + \mu'_1=\mu + \mu_1$, expressing the particle energy conservation and the particle number conservation respectively, during the collisions.

On the other hand, from relativistic kinetic theory \cite{Degroot}, it is well known that the streaming term, in the evolution equation, vanishes when the distribution function depends on the microscopic relativistic invariant, given by (\ref{B7}), which in the case of the distribution (\ref{C11}) reduced to $I=E$.

We recall that the first experimental validation of the distribution (\ref{C11}), concerns cosmic rays and has been considered in ref. \cite{PRE02}.  Recently a computer validation of the same distribution, has been considered in refs. \cite{Lapenta,Lapenta2009}, where the relaxation in relativistic plasmas under wave-particle interaction, has been simulated numerically.

As conclusions we emphasize that:

(i) The results obtained in sections II and III hold for a very large class of relativistic statistical theories, included the Juttner theory. The arbitrary function $\Lambda (f)$, after being fixed, defines univocally the statistical theory. Indeed, after fixing the form of the function $\Lambda (f)$, the two-particle correlation function defined through Eq. (\ref{B4}), the evolution equation (\ref{B1}), the stationary distribution function (\ref{B8}), and the four-vector entropy defined through Eqs. (\ref{B11}) and (\ref{H1}), are univocally fixed. Moreover during the evolution of the system the second law of the thermodynamics holds at any time, independently on the particular form of the function $\Lambda (f)$.

(ii) In fourth section, the general results obtained in sections  II and III, are employed in order to construct the kinetic foundations of the statistical theory predicting the distribution (\ref{C11}).

\end{document}